\documentclass[aps, prl,superscriptaddress, twocolumn,preprintnumbers,amsmath,amssymb]{revtex4}
\usepackage{graphicx}
\usepackage{bm}
\usepackage{amssymb, latexsym, amsmath, textcomp}

\def\br{\boldsymbol{r}}

\def\rSU2{{\rm SU}(2)}

\begin{document}
\title{Mott Insulators of Ultracold Fermionic Alkaline Earth Atoms: \\ Underconstrained Magnetism and Chiral Spin Liquid}
\author{Michael Hermele}
\affiliation{Department of Physics, University of Colorado, Boulder, Colorado 80309, USA}
\author{Victor Gurarie}
\affiliation{Department of Physics, University of Colorado, Boulder, Colorado 80309, USA}
\author{Ana Maria Rey} 
\affiliation{Department of Physics, University of Colorado, Boulder, Colorado 80309, USA}
\affiliation{JILA, University of Colorado and NIST, Boulder, Colorado, 80309, USA}
\date{\today}
\begin{abstract}
We study Mott insulators of fermionic alkaline earth atoms, described by Heisenberg spin models with enhanced ${\rm SU}(N)$ symmetry. In dramatic contrast to ${\rm SU}(2)$ magnetism, more than two spins are required to form a singlet. On the square lattice, the classical ground state is highly degenerate and magnetic order is thus unlikely. In a large-$N$ limit, we find a chiral spin liquid ground state 
with topological order and Abelian fractional statistics. We discuss its experimental detection. Chiral spin liquids with non-Abelian anyons may also be realizable with alkaline earth atoms.
\end{abstract}

\maketitle

An exciting thread in the study of strongly interacting cold atomic gases is the notion that such systems can be used as quantum simulators of strongly correlated materials \cite{jaksch05}.  Simple model systems can be engineered with a high degree of control, and studied as analogs of solid state materials.  On the other hand, in some cold atom systems the simplest realizations of strong correlation physics may have no solid state analog.  This raises the exciting prospect of systems and phenomena that are thus far unanticipated.

Recently, it has been argued that fermionic alkaline earth atoms (AEA) in optical lattice potentials can realize a variety of model correlated systems, many of which lack solid state analogs and are relatively unexplored theoretically \cite{gorshkov09}. Fermionic AEA have nuclear spins as large as $I = 9/2$ for $^{87}$Sr;  due to lack of hyperfine coupling with the electronic ground state ($^1 S_0$), the nuclear spin is essentially decoupled from the electronic degrees of freedom.  This decoupling, also present in the lowest electronic excited state ($^3 P_0$), implies that the $s$-wave scattering length is independent of nuclear spin, and leads to an enlargement of the spin rotation symmetry from ${\rm SU}(2)$ to ${\rm SU}(N)$, where $N = 2 I + 1$ \cite{gorshkov09, cazalilla09}.   This observation, together with recent progress in and prospects for manipulating AEA \cite{boyd06}, opens the door to experimental studies of ${\rm SU}(N)$ magnetism. We shall see here that the enlarged symmetry has striking physical consequences.

In this Letter, we study the simplest antiferromagnetic square-lattice ${\rm SU}(N)$ Heisenberg model that can be realized with AEA in the electronic ground state.  We find that, as in some geometrically frustrated systems, for $N \geq 3$ magnetic order is underconstrained and there is a large degeneracy of classical ground states.  Here, the degeneracy arises not from geometrical frustration but from the structure of the ${\rm SU}(N)$ exchange interaction, and is present on \emph{any} lattice for large enough $N$. This result indicates that magnetic order is unlikely, so we focus instead on non-magnetic ground states, which are controllably accessed in a large-$N$ limit, where we find the ground state is the long-sought chiral spin liquid (CSL) \cite{kalmeyer87, wen89, fradkin91, schroeter07}.  The CSL spontaneously breaks time-reversal (${\cal T}$) and parity (${\cal P}$) symmetries, and is closely related to fractional quantum Hall liquids, sharing their remarkable topological properties \cite{wen91}.  

Specifically, we consider the large-$U$ (insulating) limit of a Hubbard model with $m < N$ atoms per site.  $N \leq 10$ can be realized with $^{87}$Sr by populating a subset of the nuclear spin levels \cite{gorshkov09}.   For $m=1$, the spin at each site transforms in the fundamental representation of ${\rm SU}(N)$, and $N$ sites are needed to form a singlet, a crucial difference from ${\rm SU}(2)$ magnetism.  While $m=1$ best avoids three-body losses, we also consider $m= N/k$ for integer $k \geq 2$; in this case $k$ sites are needed to form a singlet.  Such models, which may  be realizable for $m$ not too large,  allow us to consider a solvable large-$N$ limit, where $N$ is taken large with $k$ fixed.  This is a large-$N$ generalization of the model with $m=1$ and $N = k$, as the number of sites needed to form a singlet is preserved.

It is convenient to define the model in terms of $f^\dagger_{\br \alpha}$ ($\alpha = 1 \, \dots, N$), which creates a fermion on the square lattice site $\br$.  The Hamiltonian is
\begin{equation}
\label{eqn:hamiltonian}
{\cal H} = J \sum_{\langle \br \br' \rangle} S_{\alpha \beta}(\br) S_{\beta \alpha}(\br') \text{, }S_{\alpha \beta}(\br) = f^\dagger_{\br \alpha} f^{\vphantom\dagger}_{\br \beta} \text{,}
\end{equation}
where the sum is over nearest-neighbor bonds, and $J$ is the exchange energy.  We have a local constraint, $f^\dagger_{\br \alpha} f^{\vphantom\dagger}_{\br \alpha} = m$.  Study of correction terms arising away from the large-$U$ limit will be deferred to future work.

\begin{figure}
\includegraphics[width=3.0in]{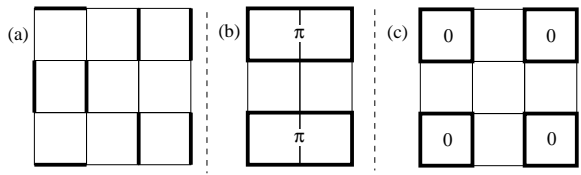}
\caption{Large-$N$ dimer and plaquette ground states for $k=2$ (a), $k=3$ (b) and $k=4$ (c).  $\chi_{\br \br'}$ has constant magnitude on the dark bonds and is zero on the others.  For $k=3$ ($k=4$), the phase of $\chi_{\br \br'}$ is chosen so that the flux through each plaquette is $\pi$ (zero).  The  patterns shown are not necessarily those selected by $1/N$ corrections.}
\label{fig:states}
\end{figure}

Most studies of ${\rm SU}(N)$ magnetism have focused on models where two sites can be combined to form a singlet.  The most-studied cases are the $k=2$ model defined above \cite{affleck88}, and models defined by placing conjugate representations on the two sublattices of a bipartite lattice \cite{read89a}.  Spin-$3/2$ alkali fermionic atoms exhibit an enlarged ${\rm SO}(5)$ symmetry, where also two sites can be combined to form a singlet  \cite{wucj03}.
Finally, we note that the models we discuss here have been solved exactly in one dimension for $m=1$ \cite{sutherland75}.  In two dimensions, the $N=4$, $m=1$ model has been studied in the context of orbitally-degenerate Mott insulators, although there the ${\rm SU}(4)$ symmetry requires substantial fine-tuning \cite{pokrovskii72}.  On the cubic lattice, plaquette states (see Fig.~\ref{fig:states}c) have been studied using a quantum plaquette model \cite{pankov07}.

\emph{Semiclassical limit.}  The semiclassical limit considered here is a generalization of the large-$S$ limit of ${\rm SU}(2)$ magnetism.  We consider a generalized model where the spin at each site transforms in the ${\rm SU}(N)$ irreducible representation labeled by the Young tableau with one row and $n_c$ columns \cite{read89a}.  This representation is the symmetric combination of $n_c$ fundamental representations, and in the ${\rm SU}(2)$ case is a spin-$S$ spin ($S = n_c/2$).  

We can define this model in terms of fermion operators $f^\dagger_{\br \alpha a}$, where $a = 1,\dots,n_c$ is a ``color'' index.  On every site we place $n_c$ fermions, and antisymmetrize over their color indices.  Defining $S_{\alpha \beta}(\br) = \sum_a  f^\dagger_{\br \alpha a} f^{\vphantom\dagger}_{\br \beta a}$, the Hamiltonian is identical in form to Eq.~(\ref{eqn:hamiltonian}).  We define the coherent state $| z \rangle = (z_{\alpha} f^\dagger_{\alpha 1}) \dots (z_{\alpha} f^\dagger_{\alpha n_c}) | 0 \rangle$, which is parametrized by the $N$-component complex spinor $z$ ($z^\dagger z^{\vphantom\dagger} = 1$) \cite{read89a}.  Since $z \to e^{i \phi} z$ only changes $|z\rangle$ by a phase, the overall phase of $z$ is unphysical and coherent states are labeled by points in the manifold ${\rm CP}^{N-1}$, which has dimension $2(N-1)$.  In the limit $n_c \to \infty$, the state $\prod_{\br} | z_{\br} \rangle$ is an eigenstate, and the energy is $E = J n_c^2 \sum_{\langle \br \br' \rangle} | z^\dagger_{\br} z^{\vphantom\dagger}_{\br'} |^2 + {\cal O}(n_c)$.

The energy is minimized for $z^\dagger_{\br} z^{\vphantom\dagger}_{\br'} = 0$ on nearest-neighbor bonds.  For $N > 2$, we immediately see a significant difference from ${\rm SU}(2)$ magnetism: knowing $z_{\br}$ does not uniquely determine the neighboring $z_{\br'}$ that minimizes the energy.  This leads to an extensive degeneracy of classical ground states.  To see this, we estimate the dimension $D$ of the ground state manifold \cite{moessner98}.  Letting $N_s$ be the number of lattice sites, the total dimension of all the ${\rm CP}^{N-1}$ spins is $2 N_s( N-1)$.  On every bond, $z^\dagger_{\br} z^{\vphantom\dagger}_{\br'} = 0$ provides two constraints, for a total of $4 N_s$ constraints.  Treating the constraints as independent leads to a \emph{lower bound}: $D \geq 2 N_s ( N - 3)$.  For $N=3$, where this bound is not helpful, it can be shown by explicit construction of ground states that $D \propto N_s$.

Such extensive degeneracy is a hallmark of geometrically frustrated systems, where underconstraint emerges from the inability to simultaneously satisfy a set of competing interactions.  A crucial physical consequence is a strong, even complete, suppression of magnetic order \cite{moessner98}.  The semiclassical limit is biased towards magnetic order, and since it is suppressed even there, we expect that the present models lack magnetic order altogether for $n_c = 1$, the case of interest for AEA Mott insulators.

\emph{Large-$N$ limit.}  Returning to the model Eq.~(\ref{eqn:hamiltonian}), magnetically disordered ground states can be be controllably studied in the limit $N \to \infty$, where $m = N / k$, $J = {\cal J} / N$, and $k$ and ${\cal J}$ are held fixed.  This limit was studied for $k=2$ in \cite{affleck88}, where the ground state is a valence-bond solid (VBS) \cite{rokhsar90}.  Here, we find the $k=3,4$ ground states break lattice symmetry and are analogous to the VBS (Fig.~\ref{fig:states}).  For $5 \leq k \leq 10$ we present evidence that the ground state is the CSL, and also discuss low-lying competing states.  We conjecture that the CSL is the ground state for all $k \geq 5$.
 
The mathematical structure of the large-$N$ limit is the same as for the $k=2$ case already studied.  The problem reduces to finding the ground state of the mean-field Hamiltonian ${\cal H}_{{\rm MFT}} = \tilde{{\cal H}}_{{\rm MFT}} + \sum_{\br} \mu_{\br} (m - f^\dagger_{\br \alpha} f^{\vphantom\dagger}_{\br \alpha} )$, where $\tilde{{\cal H}}_{{\rm MFT}}  = (N/{\cal J}) \sum_{\langle \br \br' \rangle} | \chi_{\br \br'} |^2 + {\cal H}_K$ and 
${\cal H}_K = \sum_{\langle \br \br' \rangle} (\chi_{\br \br'} f^\dagger_{\br \alpha} f^{\vphantom\dagger}_{\br' \alpha} + \text{H.c.} )$.
This is required to satisfy the self-consistency conditions
\begin{equation}
\chi_{\br \br'} = - \frac{{\cal J}}{N} \langle f^\dagger_{\br' \alpha} f^{\vphantom\dagger}_{\br \alpha} \rangle \label{eqn:sc} \text{  (a), }\qquad
m = \langle f^\dagger_{\br \alpha} f^{\vphantom\dagger}_{\br \alpha} \rangle \text{  (b).}
\end{equation}
The field $\chi_{\br \br'}$ arises from decoupling the exchange interaction on each bond, and $\mu_{\br}$ arises from a Lagrange multiplier field implementing the constraint of $m$ fermions per site.  Without loss of generality, we assume $\sum_{\br} \mu_{\br} = 0$. A set of $(\chi_{\br \br'}, \mu_{\br})$ satisfying Eq.~(\ref{eqn:sc}) is  a mean-field saddle point.  The saddle-point energy $E_{{\rm MFT}}$ is an extremum with respect to variations of the fields, but not necessarily the global minimum.  The task at hand is to find the lowest energy saddle point as a function of $k$.

For $k=2$, Rokhsar established a lower bound on $E_{{\rm MFT}}$, and showed that, on any lattice where a dimer covering is possible, any dimer state such as that shown in Fig.~\ref{fig:states}a saturates the bound \cite{rokhsar90}.  The leading corrections in the $1/N$ expansion then select an ordered VBS configuration from among the various dimer states \cite{read89a}.  It is straightforward to extend Rokhsar's bound to general $k$.  First,  for a given saddle point, using Eq.~(\ref{eqn:sc}b), $E_{{\rm MFT}} = \langle \tilde{{\cal H}}_{{\rm MFT}} \rangle \geq \tilde{E}_{{\rm MFT}}$, the ground state energy of $\tilde{{\cal H}}_{{\rm MFT}}$.  A lower bound on $\tilde{E}_{{\rm MFT}}$ is easily obtained following Ref.~\cite{rokhsar90}.  For $k=2$, one divides the spectrum of ${\cal H}_K$ in half; in general, one divides the spectrum into occupied and unoccupied levels.  On the square lattice, one finds $E_{{\rm MFT}} \geq - [ (k-1) N {\cal J} N_s] / 2k^2$.

A stricter lower bound can be established for bipartite lattices, where the spectrum of ${\cal H}_K$ is symmetric about zero energy.  We divide the spectrum into the sets ${\cal L}$ (occupied levels), ${\cal U}$ (image of ${\cal L}$ under $\epsilon \to -\epsilon$), and ${\cal M}$ (remaining levels).
An analysis similar to that of Ref.~\cite{rokhsar90} shows that, on the square lattice,
\begin{equation}
\label{eqn:bipartite-bound}
E_{{\rm MFT}} \geq - N {\cal J} N_s / 4 k\text{.}
\end{equation}
For $k > 2$ this bound is stricter than that above, and is saturated if and only if $\sum_{\br} \mu_{\br} \tilde{n}_{\br} = 0$, ${\cal L}$ (${\cal U}$) contains only the constant energy $-\epsilon$ ($+\epsilon$), and ${\cal M}$ contains only zero energy levels.  Here, $\tilde{n}_{\br}$ is the average fermion number calculated in the ground state of ${\cal H}_K$.

For $k=3,4$ the bound is saturated by the plaquette states (Fig.~\ref{fig:states}).  In each case $1/N$ corrections are expected to select an ordered configuration, and the large-$N$ ground state for $k=3,4$ is analogous to the $k=2$ VBS.  It is impossible to saturate the bound for $k > 8$:
the large-$N$ ground state energy of the two-site problem is $- N {\cal J} / k^2$, which gives the bound $E_{{\rm MFT}} \geq - 2 N {\cal J} N_s / k^2$, stricter than Eq.~(\ref{eqn:bipartite-bound}) for $k > 8$.  Even for $5 \leq k \leq 8$, the conditions needed to saturate Eq.~(\ref{eqn:bipartite-bound}) are very restrictive and we conjecture they cannot be satisfied.

Below, we present evidence that the CSL is the large-$N$ ground state for $5 \leq k \leq 10$.    The CSL saddle point has $\mu_{\br} = 0$ and $| \chi_{\br \br'} | = \chi$, with the phase of $\chi_{\br \br'}$ such that the flux through each square plaquette is $2\pi / k$.  This results in a fermion band structure  with $k$ bands, where only the lowest band is filled, and for $k \geq 3$ it is separated from the others by a gap.  This mean-field state is a lattice integer quantum Hall state:  there is a quantized Hall conductance of $N$, where the (fictitious) fermion charge and Planck's constant have been set to unity \cite{thouless82}.

\begin{figure}
\includegraphics[width=2.8in]{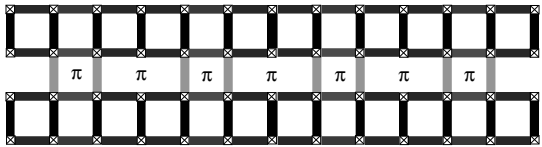}
\caption{Lowest-energy competing state for $k=6$ (the $k=5$ state has a similar pattern).  The lattice is covered by stripes, of which one is shown.  The shading of bonds represents $| \chi_{\br \br'}|$, interpolating between the maximum $|\chi_{\br \br'}|$ (black), and $| \chi_{\br \br'} | = 0$ (white).  Some regions enclose $\pi$-flux, as indicated.}
\label{fig:stripes}
\end{figure}

\begin{table}
\begin{tabular}{|r|r|r|r|}
\hline$k$ & CSL & ICSL & LC \\ \hline
5 & -0.043080 & -0.043070 & -0.042987 \\ \hline
6 & -0.033069 & -0.03299 & -0.032961  \\ \hline
7 & -0.026130 & -0.02597 & -0.025730 \\ \hline
8 & -0.021138 &  -0.02102 & -0.020897 \\ \hline
\end{tabular}
\caption{Energies of CSL and competing states, in units of $N {\cal J} N_s = N^2 J N_s$, for $5 \leq k \leq 8$.  ICSL is the lowest-energy inhomogeneous CSL that was found.  LC is the lowest competing state that cannot be interpreted as an inhomogeneous CSL.  Note that the energy difference between CSL and LC is larger for $k = 7,8$ than for $k = 5,6$.}
\label{tab}
\end{table}

To determine the lowest-energy saddle point for $5 \leq k \leq 10$, in addition to explicit construction of saddle points, we implemented a numerical self-consistent minimization (SCM) algorithm.  The algorithm begins with a random choice of $\chi_{\br \br'}$, and self-consistently iterates equation~(\ref{eqn:sc}a), while choosing $\mu_{\br}$ at each step to satisfy equation~(\ref{eqn:sc}b).  We allowed  $\chi_{\br \br'}$ and $\mu_{\br}$ to vary within a given unit cell embedded within a larger system (with periodic boundary conditions). It can be proven that SCM converges to a local minimum of the energy.  For $5 \leq k \leq 8$, we studied all rectangular unit cells with $k^2$ or fewer sites, excluding cells of unit width.  For each cell, we ran the SCM procedure on at least 30 (in some cases more than 500) different sets of random initial conditions.  The CSL was the lowest energy state found (Table~\ref{tab}).  For $k = 9,10$, less extensive application of SCM also found no states below the CSL in energy.

We also find 
\emph{locally} stable competing states, some 
 only slightly higher in energy than the CSL (Table~\ref{tab}).  The competition between CSL and these states will need to be resolved by 
   going beyond the large-$N$ limit, and, ultimately, by experiments.  The lowest such states found can be viewed as inhomogeneous versions of the CSL.  For $k=5$, a $2 \times 2$ ordering pattern is superimposed on an average $2\pi/5$ flux per plaquette.  For $k = 6,7,8$, the CSL divides into domains.  As long as the CSL remains stable to 
  inhomogeneity (\emph{e.g.} the domain wall energy is positive), these states will not be ground states.  Therefore, we also searched for the lowest competing states that \emph{cannot} be viewed as inhomogeneous CSLs.  For $k = 5,6$, we find stripe states (Fig.~\ref{fig:stripes}) that break various lattice symmetries but preserve ${\cal T}$.  For $k = 7,8$, we find a distinct CSL with $2\pi / 2k$ flux per plaquette.

\emph{Properties of CSL.}  The CSL is characterized by both its broken symmetries and topological order.  ${\cal T}$ and ${\cal P}$ breaking is signaled by a nonzero \emph{spin chirality} $\langle {\cal C}_{123} \rangle \neq 0$.  Here, ${\cal C}_{123} = i (P_{1 2 3} - P_{3 2 1})$ is the spin chirality of lattice sites $1,2,3$, and $P_{123}$ the operator that cyclically permutes the spin quantum numbers on those sites \cite{wen89}.

Understanding topological order requires going beyond the mean-field description.  It is important at this stage to note that $f^\dagger_{\br \alpha}$ does not create an atom.  Instead it creates a \emph{spinon}, which carries the spin but not the conserved atom number.  The most important fluctuations about the saddle point are in the phase of $\chi_{\br \br'} \approx  \langle \chi_{\br \br'} \rangle e^{i a_{\br \br'}}$, where $a_{\br \br'}$ is the spatial component of a fluctuating ${\rm U}(1)$ gauge field coupled to the spinons.  The time-component of the gauge field arises from the fluctuations of $\mu_{\br}$.     The gapped spinons can be integrated out to obtain a Chern-Simons (CS) effective action for the gauge field, $S_{{\rm eff}} = (N / 4\pi) \int dt d^2 \br\,  a_{\mu} \epsilon_{\mu \nu \lambda} \partial_{\nu} a_{\lambda}$, where the coefficient is determined by the mean-field Hall conductance.  The CS term is responsible for the topological properties of the CSL \cite{wen91}.  It converts spinon excitations into anyons with statistical angle $\pi + \pi / N$.  Moreover, its presence implies the spinons are deconfined and propagate freely, and the CSL thus exhibits quantum number fractionalization.  For a system with an edge, there are gapless chiral edge modes.  Finally, the ground state degeneracy on a surface of genus $g$ is $2 N^g$, where the factor of 2 arises from the spontaneous  ${\cal T}$-breaking.

\emph{Experimental detection.}  
The distinct features of the states discussed here can be split into two categories: straightforward ones associated with the spin gap and broken symmetry, and, for the CSL,  more subtle properties having to do with the presence of topological order. A number of well-developed experimental techniques can be employed to detect the features of the first type. Radio-frequency spectroscopy can be used to see the presence of the gap \cite{gupta03}. One can  measure spin-spin correlation via noise correlations to see the absence of order \cite{altman04}.  The VBS-analog states for $k= 3,4$ could be detected by adiabatically merging groups of sites into a single site, followed by application of bandmapping techniques \cite{paredes08}.  To detect ${\cal T}$-breaking, one can superimpose a second system of fermions ($^3 P_0$ alkaline earths \cite{gorshkov09}) or bosons (alkali atoms), that couples to the CSL atoms via spin-spin interaction.  By symmetry, this coupling will induce an effective orbital magnetic field for the second system.  For bosons, this field will induce vortices, and for fermions it will lead to detectable changes of the energy spectrum.  


Topological order detection is more challenging.  Due to the chiral edge modes, a disturbance of the spins near the system edge will propagate around the edge with a well-defined velocity and direction, which could potentially be detected.  Returning to the Hubbard model, the CSL will exhibit spin-charge separation.  Letting $c^\dagger_{\alpha}$ create an atom, there will be a particle carrying the atom number but not the spin, created by $b^\dagger = c^\dagger_{\alpha} f^{\vphantom\dagger}_{\alpha}$.  The $b$-particle can be thought of as a bound state of an atom and a spinon, and has fractional statistics with angle $\pi / N$.  These expectations can be formalized using a slave-rotor treatment of the Hubbard model \cite{sslee05}.  Because the spinon does not couple directly to a scalar potential,  lowering the optical potential at a lattice site can localize a $b$-particle, the fractional statistics of which could potentially be probed by techniques proposed in the context of quantum Hall-like states in cold atomic systems \cite{paredes01}. 

To observe these characteristic properties, the temperature should be at most on the order of the gap $\Delta$ to the lowest-energy quasiparticle excitations.  Using the large-$N$ limit and boldly setting $N = k$ 
(\emph{i.e.} one atom per site), in the CSL we find $\Delta \sim J$ for both excitations of the gauge field, and particle-hole excitations of the spinons. The harmonic trapping potential determines the spatial extent of $m = 1$ Mott insulator; provided this is larger than the characteristic scales of the CSL, its signatures can be observed.  
 Using the large-$N$ limit, these length scales are estimated to be at most
  a few lattice constants.

Finally, we note that the $n_c = 2$ model, discussed in the context of the semiclassical limit, can be realized using one ground state atom \emph{and} one $^3 P_0$ atom on each site, depending on the sign of the Kondo exchange \cite{gorshkov09}.  We have shown, and will present in detail elsewhere, that this model can support a CSL with a fluctuating ${\rm U}(2)$ gauge field, with a ${\rm SU}(2)$ CS term at level $N$.  This CSL supports non-Abelian anyons, and is a candidate system for universal topological quantum computation \cite{nayak08}.

In summary, we studied square lattice Mott insulators that can be realized by fermionic AEA in the electronic ground state.  We showed that magnetic order is unlikely, found the CSL ground state in a large-$N$ limit, and discussed its experimental detection.

We thank Paul Julienne for useful discussions, and acknowledge support from NIST and NSF (A.M.R.) and NSF grant no. DMR-0449521 (V.G.).

\bibliographystyle{apsrev}
\bibliography{sun-short-cr}

\end{document}